\documentclass[12pt,draftcls,onecolumn]{IEEEtran}
\usepackage{graphicx}
\usepackage{caption}
\ifCLASSINFOpdf
\else
\fi
\usepackage[cmex10]{amsmath}
\usepackage{amssymb}
\usepackage{mathrsfs}
\usepackage{verbatim}
\usepackage{graphicx}
\usepackage{epstopdf}
\usepackage{cite}
\DeclareMathOperator{\sinc}{sinc} 
\DeclareMathOperator{\rect}{rect}

\begin{document}
%
\title{Generating Waveform Families using Multi-Tone Sinusoidal Frequency Modulation}
%
%
%

\author{David~A.~Hague,~\IEEEmembership{Member,~IEEE,} \\ Naval Undersea Warfare Center \\ 1176 Howell St, Newport, RI 02840 \\ Email: david.a.hague@ieee.org}

\maketitle
\begin{abstract}
This paper presents a method for generating a family of waveforms with low in-band Auto/Cross-Correlation Function (ACF/CCF) properties using the Multi-Tone Sinusoidal Frequency Modulated (MTSFM) waveform model. The MTSFM waveform's modulation function is represented using a Fourier series expansion. The Fourier coefficients are utilized as a set of discrete parameters that can be modified to optimize the waveform family's properties.  The waveforms' ACF/CCF properties are optimized utilizing a multi-objective optimization problem.  Each objective function is weighted to place emphasis on either low ACF or CCF sidelobes.  The resulting optimized MTSFM waveforms each possess a thumbtack-like Ambiguity Function in addition to the specifically designed ACF/CCF properties.  Most importantly, the resulting MTSFM waveform families possess both ideally low Peak-to-Average Power Ratios (PAPR) and high Spectral Efficiency (SE) making them well suited for transmission on practical radar transmitters. 
\end{abstract}
\begin{keywords}
Ambiguity Function, Generalized Bessel Functions, Waveform Diversity
\end{keywords}

\IEEEpeerreviewmaketitle

\pagenumbering{gobble}

\section{Introduction}
\label{sec:intro}
%
\IEEEPARstart{W}{aveform} diversity has been a topic of great interest in the radar community for the last two decades \cite{WaveformDiversity}.  This interest has been motivated by an increasing availability of highly capable digital Arbitrary Waveform Generators (AWG) and the preeminence of cognitive and Multiple-Input Multiple-Output (MIMO) radar design concepts that can effectively utilize families of diverse waveforms.  Of particular interest are waveforms that possess a discrete set of adjustable design parameters that may be adjusted to modify waveform shape.  Waveform shape can refer to either the time-frequency characteristic of the waveform's modulation function which in turn informs its overall spectral shape, or the shape of its Ambiguity Function (AF) and its zero Doppler counterpart, the Auto Correlation Function (ACF).  These metrics for waveform shape are often utilized due to their foundational applicability to many practical systems.  Additionally, many rigorous mathematical results exist to describe their structure \cite{Cook, Rihaczek, Levanon}.  

Of the many waveform diversity problems in the literature, one that finds extensive use is the design of a family of waveforms that occupy a common operational band of frequencies and possess not only low ACF sidelobes, but also low Cross Correlation (CCF) properties with each other.  MIMO radar systems are one example application that exploits such a waveform family.  These systems transmit a unique waveform at each antenna element or sub-array to create a virtual aperture that can be much larger than standard phased-arrays utilizing a single transmit waveform \cite{jianLiII}.  This increase in virtual aperture results in enhanced parameter estimation performance and improved target resolution and localization \cite{JianLiBookI}.  A common waveform model that yields families of waveforms with desireable ACF/CCF properties are Polyphase-Coded waveforms.  A PC waveform's pulse length is divided into $N$ equal length sub-pulses known as chips. The phases of the individual chips are then assigned different phase values resulting in a distinct phase code.  The phase code represents a discrete set of adjustable waveform design parameters.  These parameters can then be modified to synthesize waveform sets which produce desired ACF/CCF properties \cite{Levanon} or other desired characterisitics which further optimize system performance.  There continues to be extensive research on designing optimal PC waveforms, specifically for MIMO applications \cite{jianLiII, PalomarIII, RangaswamyI} as well as cognitive radar applications \cite{AubryII}.  

In addition to waveform shape, there are a number of design issues to consider when transmitting waveforms on practical systems.  Perhaps the most important of these considerations is maximizing the energy transmitted into the medium to maximize detection performance in noise-limited conditions. In order to achieve this with a finite duration transmission on electronics with peak power limits, a waveform is typically required to possess a constant amplitude which translates to having a low Peak-to-Average Power Ratio (PAPR).  A constant amplitude waveform has the added benefit that it minimizes the distortion that amplitude modulated waveforms introduce to saturated power amplifiers, a common electronic component in most radar transmitters.  Additionally, it is generally desirable for waveforms to concetrate the vast majority of their energy in the frequency band of operation, a property known as Spectral Efficiency (SE).  This aids in reducing interference between systems operating in adjacent frequency bands and can also reduce perturbations in the transmitted waveform's spectral shape which may arise if the transmitter's frequency response is not ideally flat across the operational band of frequencies.  

FM waveform models generally do not possess a large set of adjustable design parameters.  However, they do naturally possess both the constant amplitude and high SE properties necessary for efficient transmission on practical devices.  PC waveforms, which do possess a sufficient number of adjustable design parameters for adaptation, have substantial spectral extent due to the instananeous phase change between chips \cite{Levanon}. This limits their overall SE and has motivated the development of Continuous Phase Modulation (CPM) techniques to improve upon their spectral characteristics \cite{Blinchikoff, BluntI}.  These CPM techniques transform the PC waveform's instantaneous phase to be continuous in its first few derivatives \cite{BluntIII}.  This effectively produces a parameterized FM waveform with constant amplitude and high SE.

Recently, the author introduced the Multi-Tone Sinusoidal Frequency Modulated (MTSFM) waveform model \cite{HagueI, Hague_Adapt} and its variants \cite{Hague_Adapt, HagueKuk} as a constant amplitude and spectrally efficient adaptive waveform design method.  The MTSFM's modulation function is represented as a finite Fourier series where the Fourier coefficients are utilized as a discrete set of parameters.  These parameters are adjusted to synthesize waveforms with specific waveform shape properties.  Recent efforts by the author have specifically explored optimizing a single MTSFM waveform's properties.  In this paper, the author explores jointly optimizing a famility of optimized MTSFM waveforms which possess both low ACF and CCF sidelobes.  This requires solving a multi-objective optimization problem involving metrics that describe the structure of the ACF and CCF.  The rest of this paper is organized as follows: Section II gives an overview of the MTSFM waveform model and the metrics used to measure its performance.  Section III demonstrates an illustrative design example which uses a multi-objective optimization problem to synthesize a family of MTSFM waveforms.  Finally, Section IV presents the paper's conclusion. 

\section{The Multi-Tone Sinusoidal FM Waveform Model}
\label{sec:format}

\subsection{The FM Waveform Model and the Ambiguity Function}
\label{subsec:AF}
The FM waveform $s\left(t\right)$ is modeled as a basebanded complex analytic signal with unit energy and duration $T$ defined over the interval $-T/2 \leq t \leq T/2$.  The waveform is expressed in the time domain as
\begin{equation}
s\left(t\right) = a\left(t\right)e^{j\varphi\left(t\right)}
\label{eq:ComplexExpo}
\end{equation}  
where $a\left(t\right)$ is a real valued and positive amplitude tapering function and $\varphi\left(t\right)$ is the phase modulation function of the waveform.  Unless otherwise specified, the amplitude tapering function $a\left(t\right)$ is a rectangular function normalized by the square root of the waveform's duration $T$ to ensure unit energy expressed as $\rect\left(t/T\right)/\sqrt{T}$.  The rectangularly windowed waveform's instantaneous frequency is solely determined by the modulation function and is expressed as 
\begin{equation}
m\left(t\right) =  \dfrac{1}{2 \pi}\dfrac{\partial \varphi \left( t\right)}{\partial t}.
\label{eq:m}
\end{equation}  
The Cross-AF (CAF) correlates a waveform's Matched Filter (MF) to the Doppler shifted versions of another waveform and is defined as \cite{Rihaczek, Cook, Levanon}
\begin{equation}
\chi_{m,n}\left(\tau, \nu\right) = \int_{-\infty}^{\infty}s_m\left(t-\frac{\tau}{2}\right)s_n^*\left(t+\frac{\tau}{2}\right)e^{j2\pi \nu t} dt
\label{eq:CAF}
\end{equation}
where $\nu$ is the doppler frequency shift expressed as $\left(\frac{2\dot{r}}{c}\right)f_c$ where $\dot{r}$ is the target's range rate, $f_c$ is the waveform's carrier frequency, and $c$ is the speed of light.  The Doppler shift is expressed in units of Hz.  The CAF simplifies to the standard Auto-AF (AAF) when $m=n$.  The waveform's ACF/CCF is the zero Doppler cut of the CAF expressed as 
\begin{equation}
R_{m,n}\left(\tau\right) = \int_{-\infty}^{\infty}s_m\left(t-\frac{\tau}{2}\right)s_n^*\left(t+\frac{\tau}{2}\right) dt
\label{eq:CCF}
\end{equation}
where the CCF simplifies to the ACF when $m=n$.

The primary focus of this paper is on waveform ACF/CCF properties and their metrics for measuring waveform performance.  The CCF measures the degree of correlation between two waveforms and provides a measure of the mutual interference between them.  Generally speaking, it is desireable for the CCF sidelobes to be as low as possible.  One metric which accurately captures this property is the area $A_{m,n}$ under $|R_{m,n}\left(\tau\right)|^2$ for all time-delays expressed as 
\begin{equation}
A_{m,n} = \int_{-T}^T |R_{m,n}\left(\tau\right)|^2 d\tau = 2\int_{0}^T |R_{m,n}\left(\tau\right)|^2 d\tau
\label{eq:CCF_Area}
\end{equation}
where the right hand side of \eqref{eq:CCF_Area} results from the even-summtery of the CCF.  A lower area $A_{\tau}$ translates to lower CCF sidelobe levels and therefore lower cross-correlation between any two waveforms.  For the ACF, there are two main design considerations.  The ACF mainlobe width determines target resolution and the ACF sidelobe structure determines the waveform's ability to distinguish a weak target in the presence of a stronger one.  Among the several definitions of mainlobe width \cite{richards2012principles}, this paper will focus on the null-to-null mainlobe width.  One particularly useful metric which provides a joint measure of mainlobe width and sidelobe structure is the Integrated Sidelobe Ratio (ISR).  The ISR is defined as the ratio of area $A_{\tau}$ under $|R\left(\tau\right)|^2$ excluding the mainlobe to the area $A_0$ under the mainlobe of $|R\left(\tau\right)|^2$ \cite{BluntI} and is expressed as 
\begin{equation}
ISR = \dfrac{A_{\tau}}{A_0} = \dfrac{2\int_{\tau_m}^{T}|R\left(\tau\right)|^2 d\tau}{\int_{-\tau_m}^{\tau_m}|R\left(\tau\right)|^2 d\tau}  = \dfrac{\int_{\tau_m}^{T}|R\left(\tau\right)|^2 d\tau}{\int_{0}^{\tau_m}|R\left(\tau\right)|^2 d\tau}
\label{eq:ISR_1}
\end{equation}
where $\tau_m$ denotes the location in time-delay of the first null of $|R\left(\tau\right)|^2$.  The mainlobe width is therefore $2\tau_m$.  As is shown in \cite{Rihaczek}, the ISR can be well approximated as 
\begin{align}
ISR &\cong \dfrac{\int_{-T}^T |R\left(\tau\right)|^2 d\tau}{\int_{-\tau_m}^{\tau_m}|R\left(\tau\right)|^2 d\tau} \IEEEnonumber\\ &= \left(\dfrac{2\beta_{rms}}{\pi}\right)\int_{-\infty}^{\infty}|S\left(f\right)|^4 df
\label{eq:ISR_2}
\end{align}
where $S\left(f\right)$ is the waveform's Fourier transform and $\beta_{rms}$ is the waveform's RMS bandwidth expressed as \cite{Rihaczek}
\begin{IEEEeqnarray}{rCl}
\beta_{rms} = 2\pi\left[\int_{-\infty}^{\infty}f^2|S\left(f\right)|^2df\right]^{1/2}.  
\label{eq:RMS}
\end{IEEEeqnarray}
The ISR components of \eqref{eq:ISR_2} are expressed as
\begin{align}
A_{\tau} &= \int_{-\infty}^{\infty}|S\left(f\right)|^4 df = \int_{-T}^{T} |R\left(\tau\right)|^2d\tau \label{eq:At}\\
A_0 &= \left(\dfrac{\pi}{2\beta_{rms}}\right) \label{eq:A0}
\end{align}
The ISR metric in \eqref{eq:ISR_2} captures the basic tradeoffs in ACF shape.  The RMS bandwidth provides a measure of the spread of the Energy Density Spectrum (EDS) $|S\left(f\right)|^2$ in frequency about DC (i.e., 0 Hz).   The mainlobe area and therefore mainlobe width of the waveform's ACF is inversely proportional to the RMS bandwidth.  The EDS follows a Parseval relation and therefore the waveform's energy is preserved in the frequency domain.  A unit energy waveform whose spectrum is tapered at higher frequencies concentrates the spread of its EDS to lower frequencies resulting in a reduced $\beta_{rms}$.  The lower $\beta_{rms}$ translates to a widened ACF mainlobe.  The $|S\left(f\right)|^4$ term in \eqref{eq:ISR_2} does not follow a Parseval relation.  In fact, a waveform whose spectrum is tapered at higher frequencies will drastically reduce the area under $|S\left(f\right)|^4$ over all frequencies.  This reduced area directly translates to a reduced area under $|R\left(\tau\right)|^2$ resulting in lower ACF sidelobes.  Reducing $A_{\tau}$ and increasing $A_0$ produces a reduced ISR which directly translates to lower ACF sidelobe levels in exchange for a widened ACF mainlobe.  For this reason, this paper uses the ISR as the primary metric to measure the ACF properties of transmit waveforms.  

\subsection{The MTSFM Waveform Model}
\label{subsec:MTSFM}

The MTSFM waveform is created by representing the modulation function \eqref{eq:m} as a Fourier Series expansion.  While a general Fourier series model will incorporate both even and odd (i.e, cosine and sine) harmonics, this paper, like previous efforts by the author, focuses on waveforms whose modulation functions possess either even or odd symmetry.  Both types of modulation functions have been analyzed in previous efforts and possess distinct resolution properties \cite{HagueIV}.  The even/odd modulation functions $m_e\left(t\right)$ and $m_o\left(t\right)$ are expressed as
\begin{align}
m_e\left(t\right) &= a_0/2 + \sum_{k=1}^K a_k \cos\left(\frac{2 \pi k t}{T}\right), \label{eq:MTSFM_e1}\\
m_o\left(t\right) &= \sum_{k=1}^K b_k \sin\left(\frac{2 \pi k t}{T}\right)
\label{eq:MTSFM_o1}.
\end{align}
Integrating with respect to time and multiplying by $2\pi$ yields the even/odd phase modulation functions of the MTSFM waveform model and are expressed as
\begin{align}
\varphi_e\left(t\right) &= \pi a_0t + \sum_{k=1}^K \alpha_k \sin\left(\frac{2 \pi k t}{T}\right), \label{eq:MTSFM_e2}\\
\varphi_o\left(t\right) &= -\sum_{k=1}^K \beta_k \cos\left(\frac{2 \pi k t}{T}\right)
\label{eq:MTSFM_o2}
\end{align}
where $\alpha_k$ and $\beta_k$ are the waveform's modulation indices expressed as $\left(\frac{a_k T}{k}\right)$ and $\left(\frac{b_k T}{k}\right)$ respectively.   Inserting \eqref{eq:MTSFM_e2} or \eqref{eq:MTSFM_o2} into the basebanded version of the waveform signal model \eqref{eq:ComplexExpo} yields the time series for a MTSFM waveform with either an even or odd symmetric modulation function respectively.  This representation for the waveform time-series does not lend itself well to deriving closed form expressions for the waveform's spectrum, AAF/CAF, or ACF/CCF.  However, there is a more convenient representation for the waveform time-series that does.  The waveform time-series can be represented as a complex Fourier series expressed as
\begin{equation}
s\left(t\right) = \frac{\rect\left(t/T\right)}{\sqrt{T}}\sum_{\ell=-\infty}^{\infty} c_{\ell} e^{\frac{j2\pi \ell t}{T}}
\end{equation}
where $c_{\ell}$ are the complex Fourier series coefficients.  These coefficients are realized as two different versions of multi-dimensional Generalized Bessel Functions (GBF) \cite{DattoliBook} with integer order $\ell$.  The type of GBF used in the expression is dependent on the symmetry of the MTSFM's modulation function \cite{Hague_Adapt}
\begin{equation}  c_{\ell} = \left\{
\begin{array}{ll}
      \mathcal{J}_{\ell}^{1:K}\left(\{\alpha_k\}\right), & m_e\left(t\right) \\
      
      \mathcal{J}_{\ell}^{1:K}\left(\{-\beta_k\}, \{-j^k\}\right), & m_o\left(t\right) \\
\end{array} 
\right.
\label{eq:SFM_Fourier_Series} 
\end{equation}
where $\mathcal{J}_{\ell}^{1:K}\left(\{\alpha_k\}\right) $ is the $K$-dimensional cylindrical GBF of the first kind and 
$\mathcal{J}_{\ell}^{1:K}\left(\{-\beta_k\}, \{-j^k\}\right)$  is the $K$-dimensional, $K-1$ parameter cylindrical GBF of the first kind \cite{DattoliBook}.  For simplicity, all closed form expressions for the MTSFM's waveform shape metrics will utilize the $K$-dimensional cylindrical GBF $\mathcal{J}_{\ell}^{1:K}\left(\{\alpha_k\}\right) $.

The spectrum of the MTSFM waveform is expressed as \cite{Hague_Adapt}
\begin{IEEEeqnarray}{rCl}
S\left(f\right) = \sqrt{T} \sum_{\ell=-\infty}^{\infty}\mathcal{J}_{\ell}^{1:K}\left(\{\alpha_k\}\right) \sinc\left[\pi T \left(f-\frac{\ell}{T}\right)\right].
\label{eq:MTSFM_Spec}
\end{IEEEeqnarray}
The AAF of the MTSFM waveform is expressed as \cite{Hague_Adapt}
\begin{multline}
\chi\left( \tau, \nu\right) = \left(\frac{T-|\tau|}{T}\right)\sum_{\ell, \ell'}\mathcal{J}_{\ell}^{1:K}\left(\{\alpha_k\}\right)  \left(\mathcal{J}_{\ell'}^{1:K}\left(\{\alpha_k\}\right) \right)^* \times \\ e^{-j\frac{\pi\left(\ell+\ell'\right) \tau}{T}} \sinc\left[\pi\left(\dfrac{T-|\tau|}{T}\right)\left(\nu T+ \left(\ell-\ell'\right)\right)\right].
\label{eq:MTSFM_NAF}
\end{multline}
The ACF of the MTSFM is obtained by setting $\nu = 0$ and is expressed as 
\begin{multline}
R\left(\tau\right) = \chi\left(\tau, \nu\right)|_{\nu=0}  \\=\left(\frac{T-|\tau|}{T}\right)\sum_{\ell, \ell'}\mathcal{J}_{\ell}^{1:K}\left(\{\alpha_k\}\right)  \left(\mathcal{J}_{\ell'}^{1:K}\left(\{\alpha_k\}\right) \right)^* \times \\ e^{-j\frac{\pi\left(\ell+\ell'\right) \tau}{T}} \sinc\left[\pi\left(\frac{T-|\tau|}{T}\right)\left(\ell-\ell'\right)\right].
\label{eq:BAF_8}
\end{multline}
The CAF and CCF between two MTSFM waveforms follows directly from \eqref{eq:MTSFM_NAF} and \eqref{eq:BAF_8} respectively.  Consider two MTSFM waveforms with modulation indices $\alpha_{1, k}$ and $\alpha_{2, k}$.  Inserting these two waveform modulation indices into the first and second GBF arguments respectively of \eqref{eq:MTSFM_NAF} and \eqref{eq:BAF_8} yields the CAF and CCF.  

The MTSFM waveform model naturally possesses a constant envelope \cite{Hague_JASA, Hague_Adapt} which satisfies the first primary requirement for transmitting waveforms on practical electronics.  Additionally, the MTSFM's modulation function is expressed as a finite Fourier series.  Any finite Fourier series is infinitely differentiable and each derivative of the modulation function is continuous and smooth \cite{boyd}.  Therefore the MTSFM's modulation function does not contain any transient-like discontinuities like PC waveforms do.  As a result of the smoothness of the MTSFM's modulation function, the vast majority of the MTSFM waveform's energy is concentrated in the waveform's swept bandwidth $\Delta f$ with very little energy residing outside of that band.  Simply stated, the MTSFM waveform naturally possesses an ideally low PAPR and high SE making them well suited for transmission on practical radar transmitter electronics.

\begin{figure}[ht]
\centering
\includegraphics[width=1.0\textwidth]{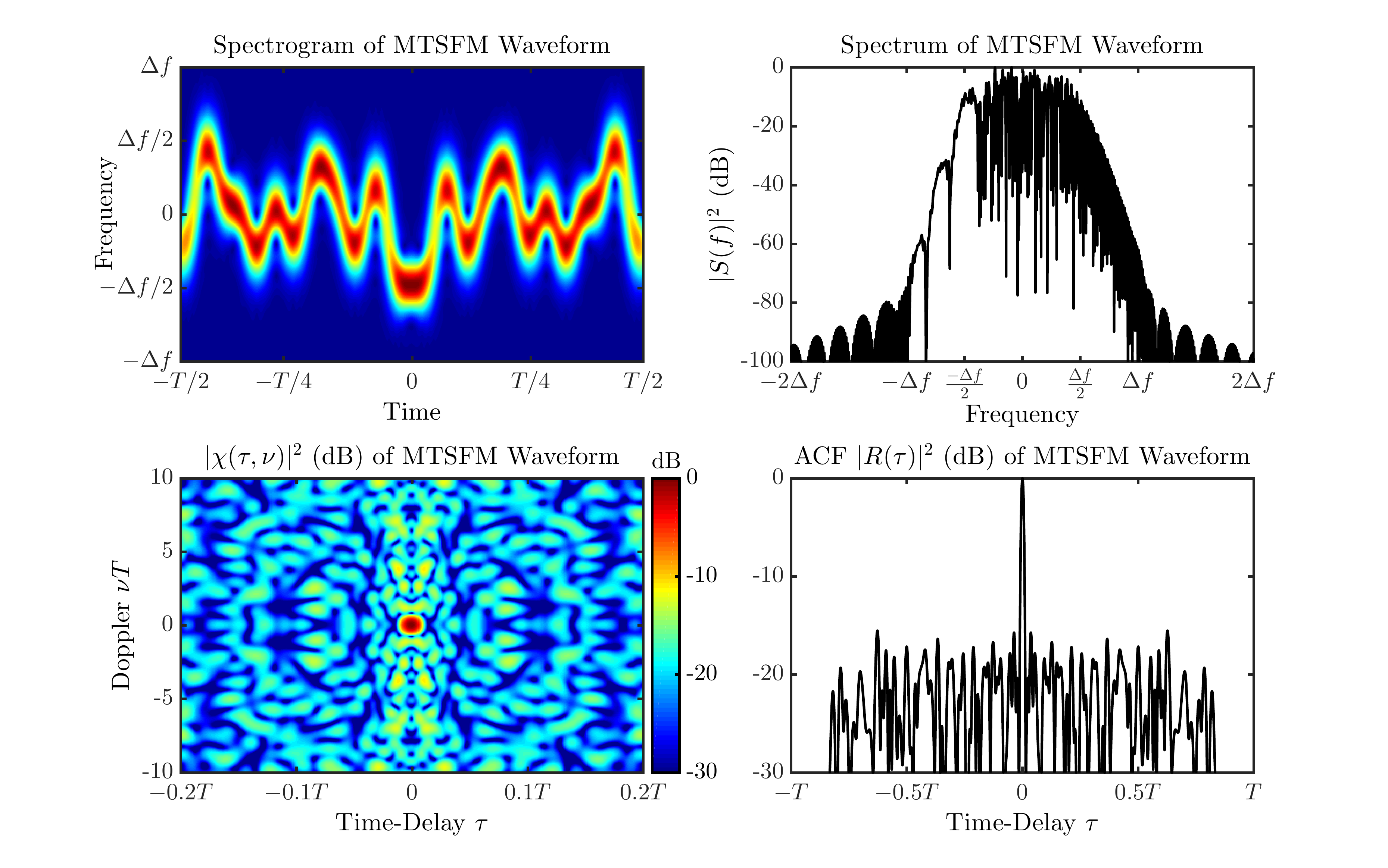}
\caption{Plot of the spectrogram (top left), EDS (top right), AF (lower left), and ACF (lower right) of an example MTSFM waveform.  The resulting waveform has a mooth modulation function resulting in a spectrally efficient waveform whose AF and ACF shapes are thumbtack-like.  The modulation indices may be modified to further refine the ACF/AF characteristics via an optimization problem.}
\label{fig:MTSFM}
\end{figure}

Figure \ref{fig:MTSFM} shows an example MTSFM waveform with a Time-Bandwidth Product (TBP) of 100.  The waveform was tapered in time using a Tukey window \cite{Harris} with shape parameter $\alpha_T = 0.05$.  This waveform's modulation function is even-symmetric as defined in \eqref{eq:MTSFM_e1} and contains $K=16$ Fourier coefficients generated using independent identically distributed (i.i.d) Gaussian random variables.  The spectrogram of the MTSFM, shown in the upper left panel, shows that its modulation function is smooth with no transient-like artifacts present.  As a result of this, the waveform's EDS, shown in the upper right panel, concentrates the vast majority of its energy in the swept bandwidth $\Delta f$ with very little energy residing outside of that band.  The pseudo-random nature of this MTSFM’s modulation function produces a waveform with a thumbtack-like AAF as seen in the lower left panel and the pedestal of sidelobes from that thumbtack-like AAF can be clearly seen in the ACF plot in the lower right panel.  The waveform in Figure \ref{fig:MTSFM} is an example of just one MTSFM waveform.  As was described in \cite{HagueI}, each new set of i.i.d random coefficients synthesizes another MTSFM waveform with a pseudo-random modulation function and a thumbtack-like AAF and possesses reasonably low cross-correlation properties with one another even when they occupy the same band of frequencies.  This paper will further refine and optimize such a family of waveforms which was not examined in \cite{HagueI}.

\section{An illustrative Design Example}
\label{sec:design}
The design of a family of waveforms with desireable low ACF/CCF sidelobes requires optimizing both the ISR of each waveform's ACF as well as the area $A_{m,n}$ under each CCF of the waveform family.  For a family of $P$ waveforms,  such an optimization problem involves modifying a set of $P$ $K$-dimensional modulation indices denoted as $\alpha_{p, k}$.  The resulting multi-objective optimization problem now must optimize over $P$ ACF ISR objective functions and $P\left(P-1\right)/2$ CCF area objective functions.  Additionally, the waveform designer may wish to place emphasis on each separate objective function via weights as the designer deems necessary.  

Formally, such a multi-objective optimization problem is defined as 
\begin{multline}
\underset{\alpha_{p, k}}{\text{min}}\left[\sum_{p=1}^P w_p\frac{ISR\left(\{\alpha_{p,k}\}\right)}{ISR\left(\{\alpha_{p,k}^{\left(0\right)}\}\right)} + \sum_{\underset{p \neq q}{p, q}} w_{p,q}\frac{A_{m,n}\left(\{\alpha_{p,k}\}\right)}{A_{m,n}\left(\{\alpha_{p,k}^{\left(0\right)}\}\right)} \right] \\ \text{s.t.~} \beta_{rms}^2\left(\{\alpha_{p,k}\}\right) \leq \left(1\pm\delta\right)\beta_{rms}^2\left(\{\alpha_{p,k}^{\left(0\right)}\}\right)
\label{eq:multiObjFunc}
\end{multline}  
where $\{\alpha_{p,k}^{\left(0\right)}\}$ are the initial MTSFM waveform coefficients fed to the optimization problem, $\delta$ is a unitless constant where $0 < \delta \leq 1$, $w_p$ and $w_{p,q}$ are weights that can emphasize and de-emphasize the individual objective functions in the optimization problem, and $\beta_{rms}^2\left(\{\alpha_{p,k}\}\right)$ is the RMS bandwidth of each waveform which is expressed as \cite{HagueI, HagueIV}
\begin{equation}
\beta_{rms}^2\left(\{\alpha_{p,k}\}\right) = \frac{2 \pi^2}{T^2}\sum_{k=1}^K \frac{k^2 \alpha_{p, k}^2}{2}.
\end{equation}
The RMS bandwidth constraint ensures the waveforms' ACF mainlobe widths do not substantially alter and thus preserves the TBP of each waveform in the waveform family.  Note that the ISR and $A_{m,n}$ objective functions are normalized by their initial values and $w_{p,q}$ are normalized such that they sum to 1.  This produces a multi-objective function whose initialized value is unity.  The resulting value $F$ of this multi-objective function after the modulation indices $\alpha_{p, k}$ are modified by the optimization routine will reside within the range $0 < F \leq 1.0$.  It is important to note that, as is described in \cite{Hague_Adapt}, the individual objective functions that compose this multi-objective function are not convex but multi-modal.  This means that the resulting value of $F$ after running this optimization problem is not  guaranteed to achieve a global minimum and that there are many local minima.  Changing the weights $w_{p,q}$ will change the location of the extrema, but not the underlying multi-modal structure of the objective function.

As a simple proof of concept of this waveform family design method, the following simulations evaluate the multi-objective optimization problem defined in \eqref{eq:multiObjFunc} for the case when $P=2$, the simplest possible case.  This problem seeks to optimize two ISR metrics and one $A_{m,n}$ metric with only three weights.  To demonstrate the variety of results achievable with the optimization problem defined in \eqref{eq:multiObjFunc}, it will be run for three different sets of weights $w_{p,q}$.  In each case, the two waveforms possess a TBP of 100 and utilize $K=64$ design coefficients initialized using i.i.d Gaussian random variables as described in \cite{HagueI}.  The waveforms are tapered in time using a Tukey window with shape parameter $\alpha_T = 0.05$.  The optimization problem is solved using MATLAB's optimization toolbox \cite{MATLAB} and $\delta = 0.2$.  

\begin{figure}[ht]
\centering
\includegraphics[width=1.0\textwidth]{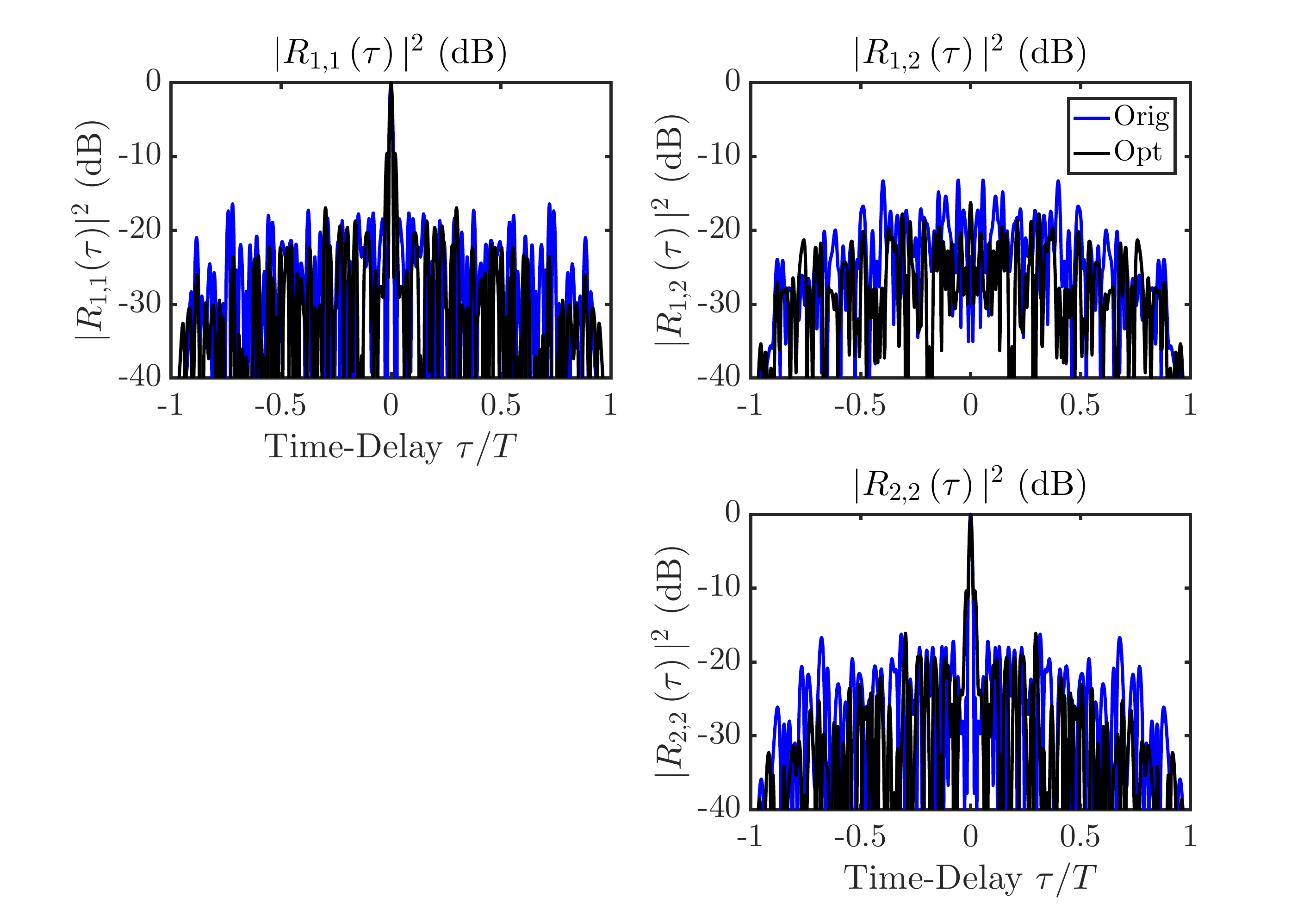}
\caption{Initialized and Optimized MTSFM waveform ACFs and CCF for two MTSFM waveforms running the optimization problem defined in \eqref{eq:multiObjFunc} with equal weighting of all objective functions.  All three objective functions were reduced slightly.}
\label{fig:equalWeight}
\end{figure}

\begin{figure}[ht]
\centering
\includegraphics[width=1.0\textwidth]{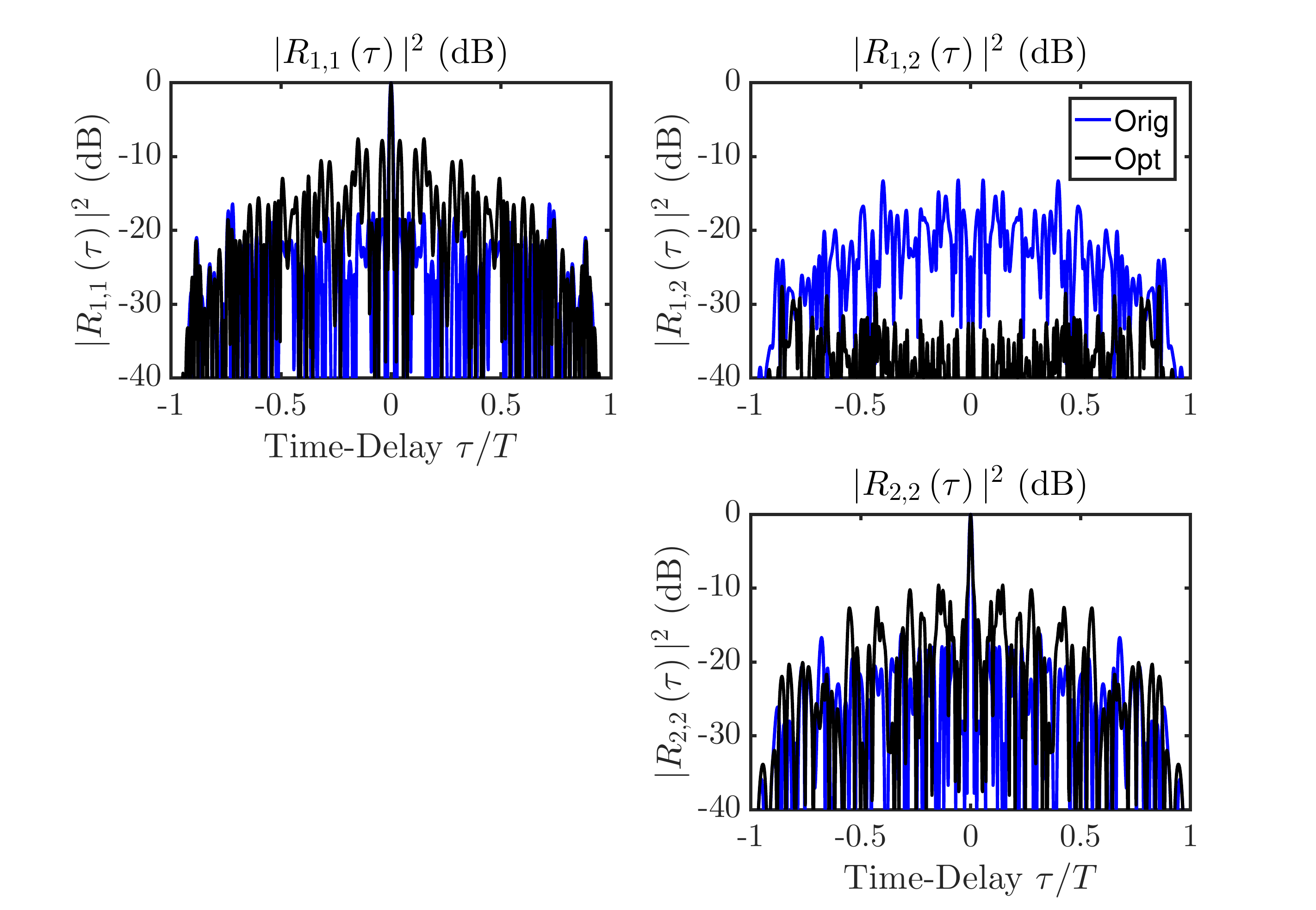}
\caption{Initialized and Optimized MTSFM waveform ACFs and CCF for two MTSFM waveforms running the optimization problem defined in \eqref{eq:multiObjFunc} where the CCF area $A_{1,2}$ objective function was weighted 10 times more than the ISR objective functions.  The resulting waveform set has noticeably low CCF sidelobe levels at the expense of increased sidelobe levels in their respective ACFs.}
\label{fig:allCCF}
\end{figure}

The first case sets all weights equal meaning each objective function is given equal emphasis in the minimization of \eqref{eq:multiObjFunc}.  The resulting waveform ACFs and CCF of this optimization run are shown in Figure \ref{fig:equalWeight}.  Overall, both waveform's ISR values were reduced slightly as was the CCF area $A_{1,2}$.  In the second case, the CCF area was weighted 10 times more heavily than the ISR objective functions.  The resulting waveforms ACFs and CCF are shown in Figure \ref{fig:allCCF}.  In this case, the CCF possesses substantially reduced sidelobes while the two ACFs possesses notably increased sidelobe levels.  This is because the CCF area was weighted so heavily that increasing the ACF ISR had little impact on the overall reduction of the multi-objective optimization problem. 

\begin{figure}[!h]
\centering
\includegraphics[width=1.0\textwidth]{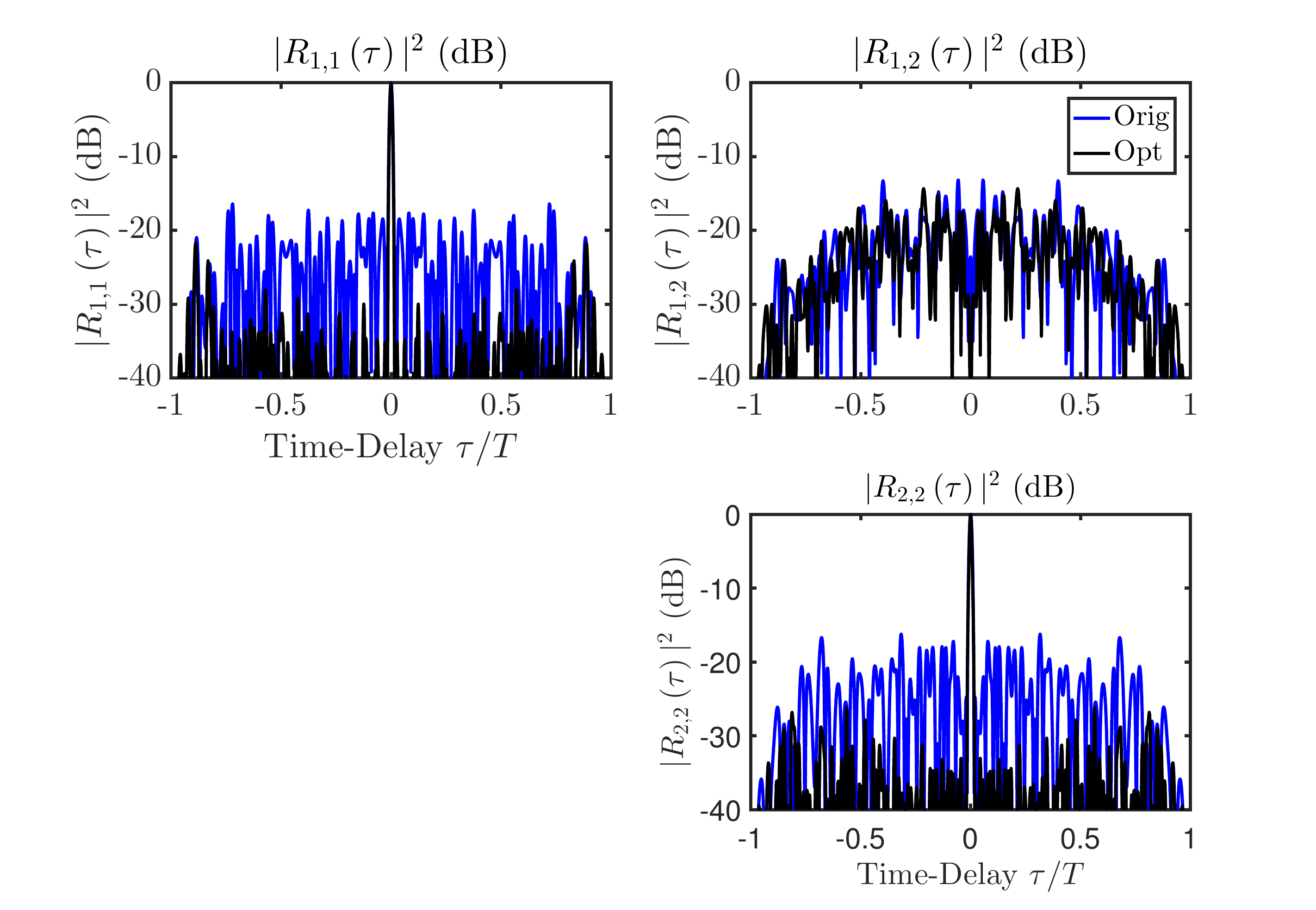}
\caption{Initialized and Optimized MTSFM waveform ACFs and CCF for two MTSFM waveforms running the optimization problem defined in \eqref{eq:multiObjFunc} where the ISR objective functions were weighted 10 times more than the CCF area $A_{1,2}$ objective function.  Surprisingly, while the ISRs for both optimized waveforms are substnatially reduced, the optimzed waveforms' CCF sidelobes remained largely fixed.}
\label{fig:allACF}
\end{figure}

In the third and final case, the ACF ISR functions were weighted 10 times more heavily than the CCF area objective function.  The resulting waveforms' ACFs and CCF are shown in Figure \ref{fig:allACF}.  The results from this case were the most surprising.  Considering that weighting the CCF heavily as in the example shown in Figure \ref{fig:allCCF} resulted in such inceased ACF sidelobe levels, one would reasonable expect that for this case the CCF sidelobes should correspondingly increase.  However, this was not the case.  As expected, both optimized MTSFM waveforms possessed substantially reduced ISRs resulting in noticeably lower ACF sidelobe levels.  However, the CCF sidelobes remained largely unchanged and in fact were reduced in some regions of time-delay.  These results demonstrate the importance the weights $w_{p,q}$ when designing families of MTSFM waveforms using the multi-objective optimization problem defined in \eqref{eq:multiObjFunc}.  Further exploration of the selection of weights $w_{p,q}$ for this and other multi-objective waveform optimization problems will be a topic of a future paper.  

\section{Conclusion}
\label{sec:Conclusion}
This paper presents a method for generating a family of waveforms with low ACF/CCF sidelobes using the Multi-Tone Sinusoidal Frequency Modulation (MTSFM) waveform model. The waveforms' ACF/CCF properties are optimized utilizing a multi-objective optimization problem where each objective function is weighted to place emphasis on either the ACF or CCF sidelobe structure.  The resulting optimized MTSFM waveforms each possess a thumbtack-like Ambiguity Function in addition to the specifically designed ACF/CCF properties.  The results from Section \ref{sec:design} demonstrate the important role the objective function weights $w_{p,q}$ play in synthesizing a family of waveforms with a collection of desireable properties.  The results presented in this paper are preliminary and only demonstrate the waveform family synthesis method for two waveforms with a specific TBP.  Future efforts will focus on evaluating this method for a collection of TBP values and for a greater number of waveforms $P$.  It is unknown at this time whether the MTSFM waveform families follow ACF/CCF sidelobe bounds in a manner similar to the Welch bound \cite{jianLiI} for PC waveforms.  Additionally, understanding the structure of the multi-objective problem defined in \eqref{eq:multiObjFunc} will likely provide insight into a more general MTSFM waveform family design methodology.  This will be the topic of a future paper.  Perhaps most importantly, the resulting MTSFM waveform families described in this paper possess both ideally low PAPR and high SE making them well suited for transmission on practical radar transmitters.

\section*{Acknowledgment}
This research was funded by the In-house Laboratory Independent Research (ILIR)  program at the Naval Undersea Warfare Center Division Newport.

\bibliographystyle{IEEEtran}
\bibliography{Hague_RadarConf20}

\end{document}